\documentclass[twocolumn]{article}
\usepackage{graphicx}
\usepackage{amsmath}

\begin{document}
\title{Yvonne Choquet-Bruhat 1923-2025}

\author{Lydia Bieri  \thanks{
    Lydia Bieri is a professor of mathematics at The University of Michigan. Her email address is lbieri@umich.edu
    } \and Jean-Pierre Bourguignon \thanks{
   Jean-Pierre Bourguignon is Nicolaas Kuiper Honorary Professor at IHES.  His email address is JPB@ihes.fr
  } \and David Garfinkle \thanks{
    David Garfinkle is a professor of physics at Oakland University. His email address is garfinkl@oakland.edu
    }\and James Isenberg\thanks{
    James Isenberg is a professor emeritus of mathematics at University of Oregon.  His email address is isenberg@uoregon.edu
   }}


\maketitle

\section*{}
\vskip -8truemm
Yvonne Bruhat was born on December 29, 1923 in Lille, France, the daughter of a philosophy professor Berthe Hubert and a physicist 
Georges Bruhat. When G. Bruhat was offered a professorship at \'Ecole Normale Sup\'erieure, the family moved to Paris, where Yvonne 
excelled both at secondary school and also at the prestigious \'Ecole Normale Sup\'erieure de Jeunes Filles. From there, she went on 
to study mathematics under the guidance of both Andr\'e Lichnerowicz and Jean Leray, holding her first scientific position at the 
French Centre National de la Recherche Scientifique, from 1949 until 1951. During the Nazi occupation of Paris during the early 
1940s, Yvonne's family suffered eviction from their home at ENS, resulting from G. Bruhat's arrest (and subsequent murder at a 
German concentration camp) for his refusal to give the Gestapo information about a student they suspected of taking part in the 
Resistance.
\vspace{2truemm}

While Einstein's theory of general relativity was formulated in 1915 in terms of geometry and partial differential equations, prior 
to Yvonne's monumental work \cite{YCB52} in 1952, most of the mathematical work involving Einstein's theory focused on generating isolated 
explicit solutions with significant symmetry. In a real sense, Yvonne introduced geometric analysis into Einstein's theory by proving 
that the Einstein equations are ``well-posed" in the sense that, for every set of initial data (consisting of a Riemannian metric $h$ 
and a symmetric tensor field $k$) specified on a three-dimensional manifold $\Sigma$, so long as that initial data satisfies the 
Einstein constraint equations, there exists a four-dimensional spacetime metric $g$ which satisfies the full system of Einstein 
equations in a spacetime close to the initial manifold. Further, $h$ is the induced metric on an embedding of $\Sigma$ in that 
spacetime, and $k$ is the induced second fundamental form. 

Yvonne's 1952 work \cite{YCB52} established  the local existence of solutions for the 
Einstein vacuum (no ``matter fields" are present) gravitational field equations. Her subsequent work showed that the Einstein field 
equations coupled to matter are also well-posed. As well, collaborating with Robert Geroch in 1969, Yvonne showed that for every 
initial data set satisfying the constraint equations, there is a unique ``maximal development" (up to spacetime diffeomorphisms): a 
spacetime solution of the Einstein equations which contains all other solutions consistent with that initial data set. Yvonne's work 
opened the path both to the serious mathematical study of astrophysical and cosmological solutions of Einstein's equations, as well 
as to the practical numerical construction of such solutions. A notable mathematical theorem regarding Einstein's equations is the 
``stability of Minkowski space" result proven by Demetrios Christodoulou and Sergiu Klainerman.

\vspace{2truemm}
The renown of Yvonne's well-posedness theorem led to a series of professorial positions at a number of universities in France, 
culminating in her 32-year academic and research career at the University Pierre-and-Marie-Curie in Paris (now a branch of Sorbonne 
University), along with her second husband Gustave Choquet. During this time, as well as following Yvonne's official retirement in 
1992, she worked on several problems, with outstanding results. Notable among these are her work on the conformal method for 
constructing and parametrizing solutions of the Einstein constraints, the proof that the Einstein equations admit gravitational wave 
solutions, the proof of the existence of global solutions of the Yang Mills equations with or without coupled Higgs fields, the 
proof of the well-posedness of the supergravity field equations, and the study of the behavior of gravitational fields in a 
neighborhood of the Big Bang singularity.

\vspace{2truemm}
Throughout Yvonne's long career, she produced several mathematical research books. Included among these are \cite{Analysis} 
(co-authored with C\'ecile DeWitt-Morette and Margaret Dillard-Bleick), and the two comprehensive 
introductions to general relativity: \cite{Choquet-Bruhat:2009xil} and the \cite{Choquet-Bruhat:2014okh} (with a Foreword by Thibault Damour). Yvonne also wrote the very 
incisive autobiographical memoir \cite{Ladymath}.

\begin{figure}
	\includegraphics[width=1.0\linewidth]{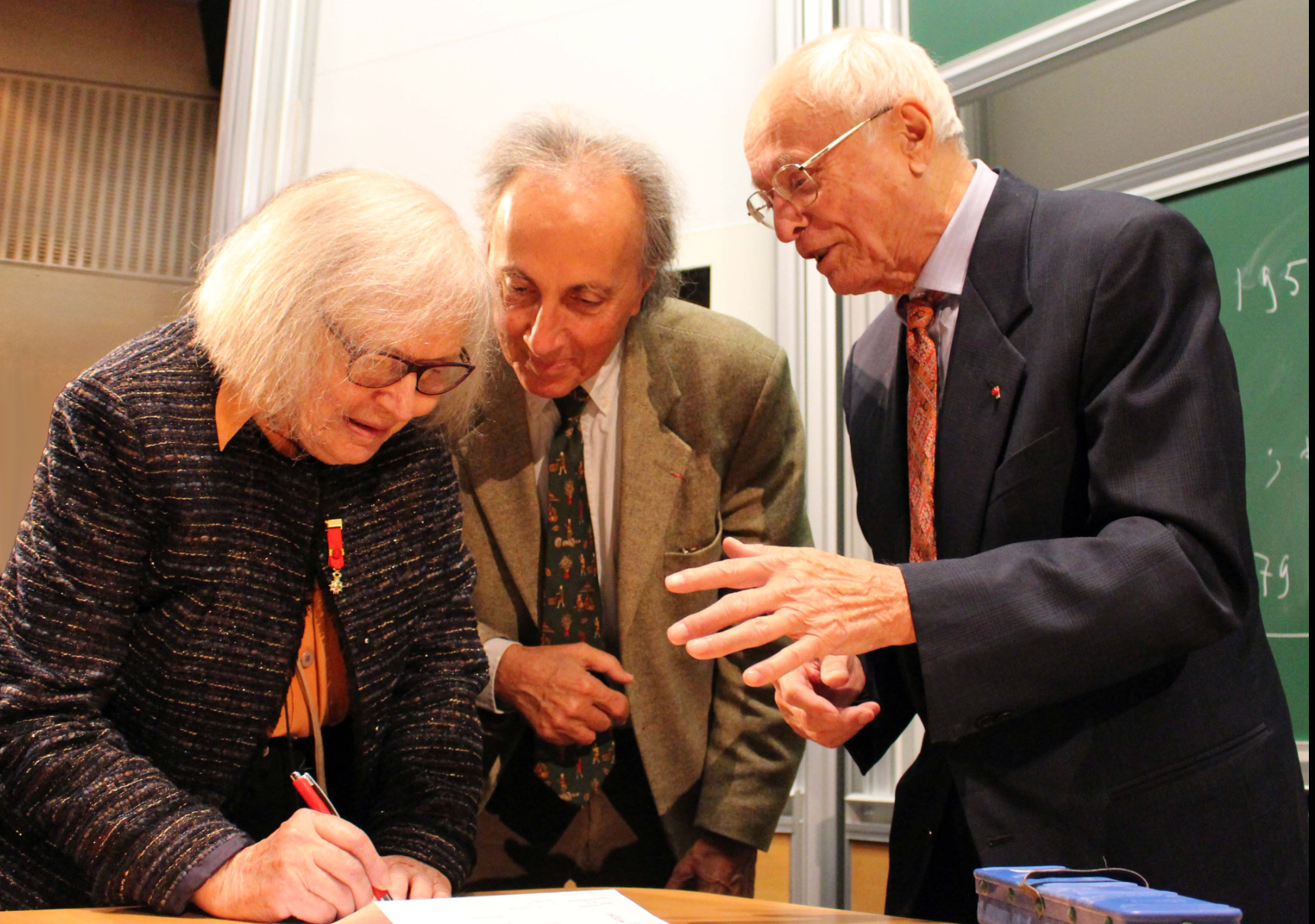}
	\caption{\small Yvonne Choquet-Bruhat signing the official document when receiving the
Grand-croix de la L\'egion d’honneur, next to Thibault Damour and Jean-Pierre Serre, her presenter 
(Photo credit IHES).}
\label{ycbfig1}
\end{figure}

Yvonne's outstanding achievements in mathematics and physics resulted in a number of prizes and honors throughout her career. Most 
notable are her elevation to a Commandeur de la L\'egion d'honneur in 1997, followed by her elevation to ``Grand Officier" and 
finally to ``Grand-croix", the highest level in the L\'egion d'honneur Order. In 2003, Yvonne was also awarded the Dannie Heineman Prize for 
Mathematical Physics by the American Physical Society.
\vspace{2truemm}

After her retirement in 1992, Yvonne continued doing research for many years, working primarily at the Institut des Hautes \'Etudes 
Scientifiques. 

Yvonne died on February 11, 2025 (at the age of 101) in M\'erignac, France, surrounded by her family.

\subsection*{Abhay Ashtekar}

It is a pleasure and great honor to participate in this celebration of Madame Choquet's towering lifetime achievements. She has been 
an inspiring figure for me since I studied her early work on Einstein's equations as a Ph.D. student of Bob Geroch's 50 years ago. 
Already then she was a legend as an early pioneer, exploring these equations using rigorous methods from geometric analysis. Indeed, 
papers she wrote in the 1950s are such lofty landmarks that they continue to be important, and cited even today! 

My admiration for her work grew over the subsequent years as I came to know more about her deep contributions, not only to general 
relativity but also to the global existence and uniqueness of solutions to gauge theories. I then had the privilege of being her 
colleague at the Jussieu campus in the eighties, as a {\it Professeur} at Paris VI. Our paths did not cross regularly. But her warm 
welcome in Paris, and scientific discussions at seminars and  \textit{``Journ\'ees Relativistes"}, were among the highlights of my academic 
life in France. The Paris academic scene was not easy for me to navigate and I am grateful to have had a guide in Madame Choquet 
during those years.

In the subsequent 20 years or so my own research was focused more on quantum gravity and even my contributions to classical 
general relativity have been more at the interface of geometry and physics rather than geometry and analysis. Nonetheless, I always 
found Madame Choquet's work inspiring. Her book, {\it Geometry, Analysis and Physics} has been among the most used volumes on the 
book-shelf in my office since late 70's! When she was awarded the Dannie Heineman Prize by the American Physical Society in 2003, I 
wrote to her expressing my sentiments on hearing the news: Most prizes honor the recipients, but on rare occasions it is the 
recipients who make the prize more prestigious. And I do feel that the Dannie Heineman Prize became much more significant in the 
general relativity circles after 2003. 

In more recent years, I had the privilege of interacting with Madame Choquet in person at a 
KITP workshop in Santa Barbara, a Marcel Grossmann Meeting in Rome and at the celebration of the 100$^{\textrm{th}}$ anniversary of General 
Relativity in Berlin. On the Santa Barbara campus, we all watched with great admiration as she rode her bicycle in a cheerful, carefree manner. She had not changed over all those years -- there was the same kindness radiating from her, the same mathematical rigor, 
and the same simplicity in her demeanor, providing us with a role model on how to live a productive and fulfilling life with inner 
serenity.

And then, over the last couple of years, I came to read of her life as a student in occupied France. In 1944, when Yvonne's  father, 
Georges Bruhat -- a physicist and the Deputy Director of the Paris \'Ecole Normale Sup\'erieure-- refused to cooperate with the 
Gestapo, they arrested Berthe Hubert Bruhat -- Yvonne's mother -- and threatened to kill her the following day. Yvonne went to the 
headquarters and pleaded with the Gestapo to allow her to take her mother's place. When I read this, my heart melted and there were 
tears in my eyes. This was a whole new dimension of Madame Choquet that I knew nothing about. 

To have made such deep contributions to 
mathematics {\it and} to have attained the inner serenity in spite of enduring such traumatic events early on, made my already 
enormous admiration and respect grow a thousand fold. What a remarkable life you led, Madame Choquet! I feel fortunate to have known 
you in person.

\subsection*{Lydia Bieri}
The first time I met Yvonne was in 2005 at a conference in Cambridge (GB). During that time, I was a Ph.D. student at ETH in Zurich, 
and I had read many of her publications. Even before meeting her, I was deeply impressed by both the profound depth and remarkable 
breadth of her work. Meeting her was both impressive and refreshing. Yvonne was open-minded, sharp, direct, and kind, always 
welcoming to students and junior colleagues. Over the years, our ways kept crossing, including during my trips to Paris. And I recall 
delightful discussions with Yvonne during one of my stays at the Institut des Hautes \'Etudes Scientifiques (IHES) in 2016. She 
regularly came to the Institute and truly enjoyed discussing mathematics and physics with her friends and colleagues. During lunch, 
these conversations often expanded to cover life and society. Yvonne's profound intellectual curiosity spanned many subjects. Not 
only had she explored the intricate structures of the universe, but she had also reflected about the relationship between humans, 
nature and life. Discussing with her it became clear that her work was driven by a deep ambition to comprehend the universe through 
mathematics and physics. 

Yvonne explored, pioneered and established significant insights into the interweaving of mathematics and physics in general 
relativity, thereby opening up the field to new avenues of mathematical exploration. 
A fantastic example is her early work (1952) on the Cauchy problem marking the beginning of the mathematically rigorous approach to 
the Einstein equations. There, Yvonne proved a local existence and uniqueness theorem for the Einstein equations. By means in this 
proof, the Einstein vacuum equations take the form of a system of hyperbolic nonlinear (quasilinear) partial differential equations. 
This result also established the first proof that for the nonlinear Einstein equations gravitational waves propagate at finite speed. 

For decades prior, a significant degree of uncertainty and debate surrounded this very concept within the physics community. This 
stemmed from Einstein's early derivations of wave solutions in 1916 in the framework of the linearized equations, which raised 
questions about whether the full theory, that is the nonlinear Einstein equations, allow for gravitational waves or not. Yvonne's 
proof resolved this issue, answering with a clear yes. 

In connection with this, Yvonne recounted the following episode (that she also shared in her memoirs) describing how she met Albert 
Einstein for the first time at the Institute for Advanced Study in Princeton, where she had arrived in 1951 for a postdoc position. 
In their first meeting, Einstein asked Madame Bruhat to explain her work to him. The following are Yvonne's words, citing from her 
memoirs: \textit{`` .... I summarized, in French, my thesis on Einstein's blackboard. Einstein listened with interest, and congratulated me 
for the results which gave rigorous proofs of properties he had expected from the gravitational field. When I left, Einstein
told me I would be welcome in his office any time I felt like knocking on his door ...."}
Later on, Yvonne continued to contribute extensively to gravitational wave research, in particular through her work on the theory of 
hyperbolic partial differential equations. 

Yvonne will be deeply missed. We will continue to remember her not only for her brilliant mind but also for the compassionate spirit 
that inspired many. 

\subsection*{Jean-Pierre Bourguignon}

From 2003 to 2015, Yvonne Choquet-Bruhat offered IH\'ES the privilege of regular visits, well after she took emeritus status 
at her University. Of course this gave her the opportunity of discussing extensively with Thibault Damour, a permanent 
professor of the Institute, but also with a number of visitors, some of them long term collaborators of hers. I myself did also take 
advantage scientifically of her regular presence, in relation with my teaching (for 15 years) the mathematical side of General 
Relativity at \'Ecole polytechnique, jointly with, successively, Nathalie Deruelle and David Langlois. 

Several contributors to this homage, who directly collaborated with her, will focus their input on her major contributions to the 
Theory of General Relativity, an extraordinary lifetime achievement. For this testimony, I would like to focus on two aspects of her 
legacy: her books, and her remarkable talent for using the right words on public occasions. 

Still, I cannot begin this note without briefly recalling the extraordinary breakthrough that the results she obtained in her 1952 
thesis \cite{YCB52} represented: she provided the first ever general solution of the system of Einstein equations, yet local in space 
and time; these results also provided mathematical evidence for the existence of gravitational waves. With time passing and the 
technical development of the theory of (systems of) partial differential equations making considerable progress in the second half of 
the 20$^{\rm th}$ century, in particular dealing with many non-linear situations, we might underestimate the phenomenal achievement 
Yvonne made then.

Let me now discuss the books which she wrote. In the Epilogue (of the 351-page English translation) of her 
autobiography \cite{Ladymath}, she states: 
{\it ``One can see in my biography that I had had a full life, three children, seven books, some three hundred articles and many 
friends"}. Indeed, all along her career, and even long after retiring from the University, she wrote very comprehensive and epoch-making books. Here, I would like to concentrate my comments on three of them: 
\begin{itemize}
\item
\vskip -3truemm
the first book I note is \cite{Analysis}, in two volumes published in 1977 and 1989, was written with her friend C\'ecile DeWitt-
Morette (and also with Margaret Dillard-Bleick for the first volume); the material presented in it mixes in an efficient 
way Analysis, Geometry and Theoretical Physics, an approach which, at the time, was highly original and completely in phase with the 
new research developments in Global Analysis occurring then; in lectures I gave at the time, I personally made extensive reference 
to this book. It is quite remarkable that the first edition was reprinted 7 times with several revisions;
\item
\vskip -3truemm 
the second book which I note is the impressive 785-page long volume \cite{Choquet-Bruhat:2009xil}, completed when 
she was already 86 years old; in it, one finds all the foundations but also a remarkable up-to-date presentation of the most recent 
research results including those dealing with global solutions of the Einstein equations;
\vskip -3truemm
\item
\vskip -3truemm
the third book I point out is \cite{Choquet-Bruhat:2014okh}, in some sense even more impressive; it was  published in 2015 -- Yvonne was 
then 88 -- which is a masterpiece of pedagogy while being comprehensive; here is how Thibault Damour phrases his 
appreciation in the Foreword to the book: {\it ``She has succeeded in reaching a Landau- and Lifshitz-like ideal of covering all the 
crucial issues in the most concise way, while expounding each topic in a mathematically rigorous way. This rare combination of 
qualities makes this book particularly valuable"}.
\end{itemize}

I now come to her remarkable mastery in expressing herself in texts or in public. Many examples could be given. I give two from 
occasions I directly witnessed:
\begin{itemize}
\item
\vskip -3truemm
I first quote the speech which she gave on the occasion of Chern Shiing Shen's ${100}^{\rm th}$ birthday
celebration at IHES; her conclusion was: {\it ``Des personnalit\'es telles que [la sienne], d'une remarquable intelligence, noblesse 
et g\'en\'e\-rosit\'e redonnent un peu confiance dans l'avenir de l'humanit\'e dont la Chine constitue actuellement le 
quart"}.\footnote{\it ``Personalities such as [him], with remarkable intelligence, nobility and generosity, restore some confidence 
in the future of humanity, a quarter of which currently resides in China."} ;
\item
\vskip -3truemm
The second presentation which I cite is her dedication speech when her close friend C\'ecile DeWitt-Morette chose her as 
presenter when receiving the second degree of the French L\'egion d'honneur; I provide just one quote from there: {\it ``Par 
contre C\'ecile..., int\'eress\'ee \`a la g\'eom\'etrie diff\'erentielle moderne \`a l'\'epoque, a m\^eme offert de traduire en 
anglais un livre relativement \'el\'ementaire que j'avais \'ecrit sur ce sujet. Je trouvais que c'\'etait indigne d'elle et je lui 
ai sugg\'er\'e que nous \'ecrivions  ensemble un livre plus complet contenant aussi de l'analyse et des applications \`a la 
physique, qui pourrait servir aux physiciens"}\footnote{\it ``Nevertheless, C\'ecile...., interested then in modern differential 
geometry, even offered to translate into English a relatively elementary book I had written on the subject. I found that this was  
unworthy of her and I suggested to her that we write together a more complete book covering also analysis and applications to  
physics, which could be useful to physicists."}; this provides a hint on the genesis of the book \cite{Analysis} I quoted and praised earlier.
\end{itemize}

\begin{figure}
	\includegraphics[width=1.0\linewidth]{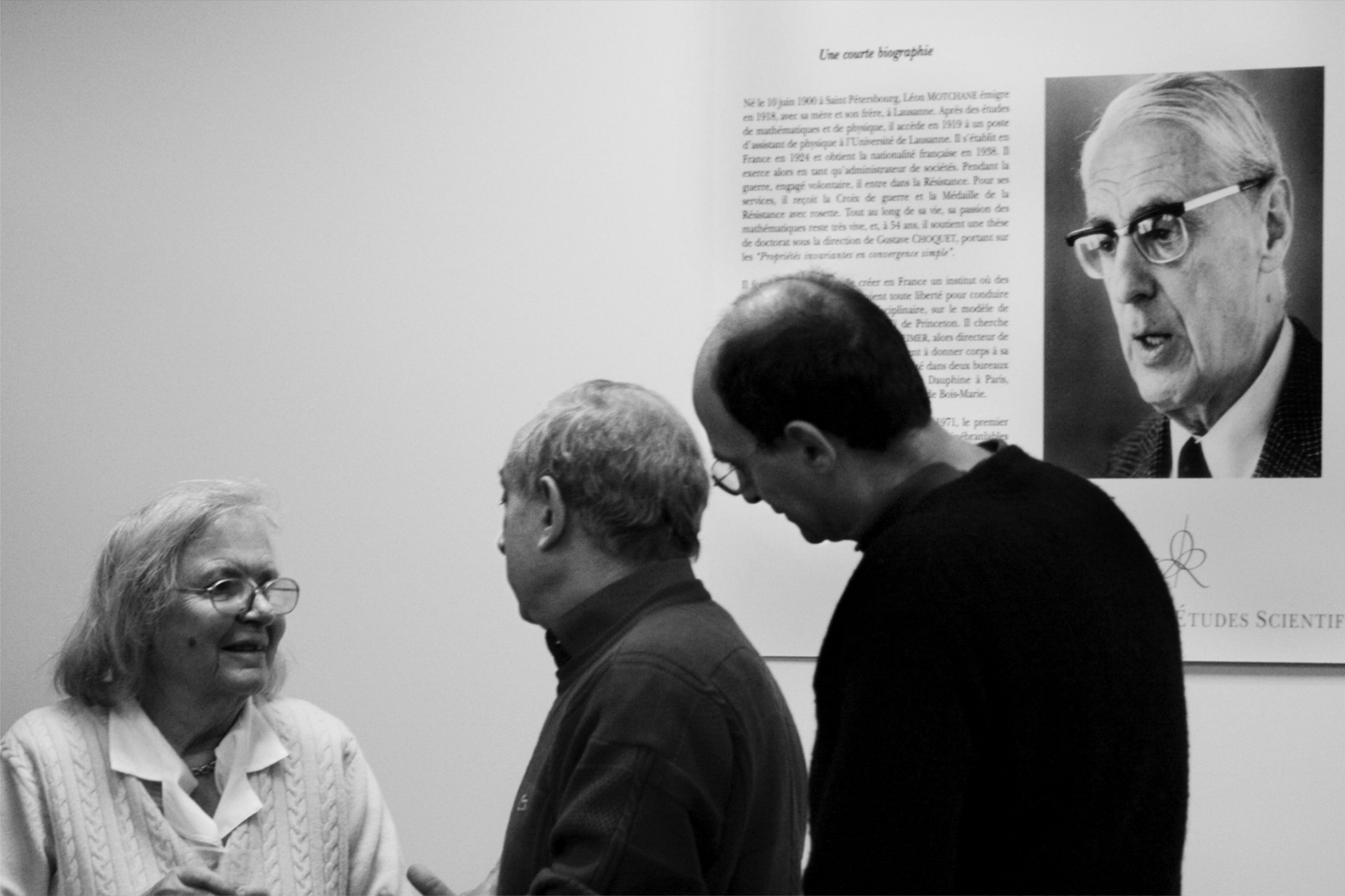}
	\caption{\small Yvonne Choquet-Bruhat discussing with Claude Zuily and Sergiu Klainerman
in front of the portrait of L\'eon Motchane at IHES  (Photo credit Jean-Fran{\c c}ois Dars).}	
\label{ycbfig2}
\vspace{-5truemm}
\end{figure}

The text I want to end with is an excerpt from Yvonne's contribution to \cite{unravelers} entitled 
{\it ``Knowing, understanding, discovering"}. It is so perfectly written that it is very hard to 
take one word out of almost any paragraph without altering badly the ideas she expresses. Here is how she ends, once again stressing 
the importance of ``friendship" for her: {\it ``... the mathematical universe exists through the community of mathematicians who 
create it -- or who discover it if the reader prefers that philosophy. It is a great joy for a mathematician to belong to this 
community of citizens of the same ideal country. Specialists of the same discipline, whatever their nationality, share a certain 
number of truths and curiosity for unresolved problems. Their common knowledge and interests unite them more than rivalries of 
priority could divide them. Exchanges of points of view are stimulating and enriching. Work in collaboration is particularly 
gratifying. Mathematical affinities sometimes turn into real friendship, the salt of life."}

\subsection*{Demetrios Christodoulou}
I first met Yvonne Choquet-Bruhat in Turin in late fall of 1977, where we were both guests of the late Mauro Francaviglia. Prior to that year 
I had acquired no further knowledge of mathematics than that of the average physics
student. However earlier that year J\"urgen Ehlers, who was the head of the group at the Max Planck Institute in Munich where I was working as 
a postdoctoral fellow, had realized that I had an aptitude for mathematics and had
encouraged me to first study ordinary differential equations and then to move on to partial differential equations under the guidance of 
Yvonne Choquet-Bruhat in Paris. She had established in 1952 the fundamental local in time
existence theorem for the Einstein equations of general relativity. Thus I took the opportunity to meet her for the first time in Turin. She 
invited me to Paris in the early spring of 1978, and this was for me the beginning of a 4 year learning experience, spending most of that time 
in Paris, learning basic concepts and methods, from energy estimates to Leray-Schauder degree theory.
It was during this time that I had the honor of interacting with Jean Leray who was Yvonne Choquet’s close friend and mentor.

During one of Yvonne Choquet’s visits to Germany while I was still at the Max Planck Institute an amusing incident occurred. J\"urgen Ehlers 
had invited both of us for dinner at his mansion on a lake south of Munich. At the dinner,
plenty of excellent wine was served. And J\"urgen Ehlers, experiencing a touch of joviality, at one point turned to Yvonne Choquet and asked 
her whether she had been a student of Henri Poincar\'e. To which she replied, visibly annoyed, “What do you think I am, ancient?”. In fact, 
Yvonne Choquet was born more than 11 years after Poincar\'e passed away and she was to outlive Ehlers by more than 16 years.

J\"urgen Ehlers and Yvonne Choquet secured for me the Max Planck Medal, and with the funds from this award I moved back to the U.S. in the 
Fall of 1981. I had been in the U.S. in the period 1968-1972, obtaining my Ph.D. in physics from Princeton in 1971, but this time I joined a 
different community, the mathematical community. I had to return to Europe in the summer of 1982 to obtain an immigrant visa for the U.S. 
where my visiting membership at the Courant Institute was to start in September. At the end of the summer, I visited Yvonne Choquet in Paris. 
However, my visa was not ready when September came and my documents were transferred to the U.S. Embassy in Paris where
I stayed for two more months hosted by Yvonne Choquet, a memorable and productive period for me. Returning to the U.S. in late fall of 1982  I 
learned geometric analysis from Shing-Tung Yau, this completing my mathematical education.

 I met Yvonne Choquet several times after that, notably on the occasion of a conference in Poland in 2006 in memory of Leray and Schauder, 
 during my month-long visit to the Coll\`ege de France in May 2009, and lastly during a visit to Paris in November 2015, where I was present 
 in a ceremony at IHES when she was elevated to the Grand Croix de l’Ordre national de la L\'egion d’honneur (Grand Cross of the National 
 Order of the Legion of Honor), the highest echelon of the L\'egion d’honneur and the highest recognition in France limited to 75 persons. On 
 that occasion Jean-Pierre Serre, who gave the presentation speech, to illustrate her work on the Einstein equations, said that she proved 
 that the Cauchy problem is well-posed while Andrew Wiles proved that the Fermat problem is ill-posed. 

\subsection*{Thibault Damour}

\noindent
\textbf{First encounter}
 
My earliest memory of Yvonne Choquet-Bruhat (whom I will simply call Yvonne) is her smiling face when we first met on a train platform in 
Paris. Was it July 1975, on our way to the first Marcel Grossmann Meeting in Trieste? I don't know, but I do clearly remember her warm, 
welcoming smile toward a junior scientist just starting out in gravitational physics. That smile never faded throughout all the years 
I had the pleasure of interacting with her -- particularly between 2003 and 2018, when she was a regular visitor at IHES.

\subsubsection*{First interactions linked to science}
 
The Comptes Rendus of the (French) Academy of Sciences were originally created to announce, briefly and rapidly, significant new 
results. In particular, they allowed junior scientists to publish a short account of novel results under the aegis of a member of the 
Academy of Sciences who served as a ``referee", vouching for the (a priori) validity of their results.
For instance, Yvonne's first announcement of her breakthrough result on  Einstein's field equations, at the early age of 26,
namely her \textit{``Th\'{e}or\`{e}me d'existence pour les \'{e}quations de la gravitation einsteinienne dans le cas  non analytique"}, was 
published in early 1950, a couple of weeks after the presentation of her Note to the Academy, on February 6, 1950,  by Jacques Hadamard 
(one of the greatest living mathematicians at the time). See \cite{YCB50} for an English translation, and republication, of  Yvonne's Note 
as a Golden Oldie, and \cite{DCRK2022} for a beautiful accompanying editorial comment.

I had become aware of this possibility of ``recording the date" (``prendre date", in the words of Lichnerowicz)
 in the Fall of 1974  when Remo Ruffini introduced me to Andr\'e Lichnerowicz
in Princeton. [And this eventually led to the fast publication (December 1974) of my first published paper, a Comptes Rendus 
Note with Remo on the newly discovered Hulse-Taylor binary pulsar \cite{DamourRuffini}.] 
The reason for my mentioning this way of helping junior scientists to publish rapidly novel results (and also to bring their
work to the attention of senior experts in their field) is that, around 1980, when Nathalie Deruelle and I obtained
novel results on the two body problem in General Relativity we asked Yvonne to present to the Academy two
short Notes \cite{DD81a,DD81b} on the Lagrangian dynamics of two point masses at the second post-Newtonian approximation,
soon followed by another Note by me \cite{D82} also presented by Yvonne.
Nathalie and I were quite grateful to Yvonne for this triple opportunity of expeditiously announcing our new results
at a moment where there was an international, competitive effort to clarify the so-called ``quadrupole 
formula controversy", i.e., the issue of understanding the back reaction of gravitational radiation on the
dynamics of binary systems, such as the Hulse-Taylor binary pulsar.

 \subsubsection*{Deeper scientific interactions with Yvonne}
 
 Though, as mentioned above, I had met Yvonne early in my career, and I continued meeting her,
 in an increasingly friendly way, in many subsequent scientific meetings (Marcel Grossmann meetings, Journ\'ees 
 Relativistes\footnote{Let me mention in this respect that Yvonne reminded me a few times
 that, though many people attributed the conception (around 1968) of those yearly French relativistic days
 to Lichnerowicz (who was happy of this attribution), the idea originally came from her, or was at least conceived
 in consultation between her and Lichnerowicz.}, GRG meetings, etc.), I must confess that, for many years,
 I had no real scientific interactions with her. Seen from the perspective of a junior theoretical physicist,
 interested in exploring the observable consequences of Einstein's gravitation theory 
 in our real Universe, most of her work appeared to me as being too mathematically-oriented to 
 attract more than a rather far-away contemplation. It took me years to understand that I was wrong,
 and that Yvonne's works contained many results of direct importance for gravitational physics.
 
 Let me  mention just a few of the physics-relevant results of Yvonne (besides her pioneering
 non-analytic existence theorem \cite{YCB50,YCB52} whose physics importance I also only slowly appreciated,
 and her many results on the Einstein constraint equations, which, I assume, are discussed in other contributions to this homage):
 
 \begin{itemize}
 
\item  First, her pioneering work \cite{FB56} on the $3+1$ decomposition of Einstein's equations, usually attributed only
to Arnowitt-Deser-Misner\footnote{Let me mention in this respect that Yvonne once told me (I think that she wrote this in her autobiography)
that Stanley Deser (a good friend of both of us) had explicitly told her that there was no need of her including
a discussion of the result \cite{FB56} in her contribution to Louis Witten's famous book, \textsl{Gravitation: an Introduction 
to Current Research} because \textit{``it was well-known".}}. As we know, the $3+1$ decomposition of Einsteinian gravity has become
one of the key elements of modern Numerical Relativity. I present more about this below.
 
\item  Second, her work on strong high-frequency gravitational waves \cite{YCB69}. It took me time to grasp what
made this work significantly superior to the famous, slightly earlier, work of Rich Isaacson \cite{Isaacson:1968zza}.
Besides a mathematically clear way of defining (by the two-timing technique) and justifying (by a boundedness condition)
the averaging leading to the appearance of the effective stress-energy tensor of gravitational waves,
the most interesting result of \cite{YCB69} is the realization that the nonlinear structure of Einstein's equations
ensures the lack of steepening of strong gravitational waves during their propagation. In other words, nonlinear
gravitational waves satisfy the exceptionality condition of Lax and Boillat (see, e.g., \cite{Boillat:1969slt}
and references therein.) This exceptional property of nonlinear gravitational waves is directly related to the weak form of the
Christodoulou-Klainerman ``null condition" satisfied by Einstein's equations (see \cite{Lindblad03} and references
therein.)  For recent progress in the construction of exact solutions containing strong high-frequency gravitational
waves see \cite{Touati:2024xpw} and references therein.
 
\item  
Then I would like to mention Yvonne's work on the positive mass issue. Most importantly, her 1976 work
with Jerrold Marsden \cite{YCB76} gave the first rigourous proof of the positivity of energy for vacuum 
space times near flat space using a critical point analysis in infinite dimensions.  See below for a later
contribution of Yvonne, and \cite{Huang25} for a recent survey of positive(-mass)-energy theorems.
 
\item 
Of particular importance is the work of Yvonne on hyperbolic formulations of 3+1 versions
of the evolution part of Einstein's equations.  In particular, Yvonne and Tommaso Ruggeri obtained
(by combining the evolution equations with the constraints) a 3+1 hyperbolic system with zero shift $\beta$.

This was then generalized by Yvonne (in collaboration with  Jimmy York and Arlen Anderson) to the construction
of hyperbolic 3+1 systems with non-zero shift. These results were part of an international effort to construct
3+1 evolution systems having sufficient stability to be solved on a computer. 

Let me note in particular that the
crucial time-hyperbolicity condition used by Yvonne and Tommaso Ruggeri, which yields, in 3+1 variables,
the evolution equation for the lapse $\alpha$, $(\partial_t - {\mathcal L}_{\beta}) \alpha = - \alpha^2 K$,
was later generalized   \cite{Anninos:1995am} in the so-called ``1+log" slicing condition, 
$(\partial_t - {\mathcal L}_{\beta}) \alpha = - 2 \alpha K$ which has become an important ingredient of many
modern Numerical Relativity codes.
 
\item  Let me finally mention the important contributions of Yvonne to understanding the
causality properties of several of the extensions of Einstein's theory suggested by modern theoretical physics.
I have notably in mind here the works of Yvonne on Supergravity \cite{Choquet-Bruhat:1983xyr,Bao:1984bp,Choquet-Bruhat:1985xei} 
and on Gauss-Bonnet-type gravity \cite{Choquet-Bruhat:1988jdt}.

\end{itemize}
  
\subsubsection*{Discussions with Yvonne at IHES}
  
From 2003 to 2018, Yvonne regularly visited  IHES (at least once per week), 
and I had the pleasure to discuss many times with her, about either mathematics or physics. This also allowed me to have many enlightening discussions 
with several of the collaborators of Yvonne who visited IHES in those years, notably Vince Moncrief and Piotr Chrusciel. 

Let me first recall that I was always amazed by the clearness and accuracy of her answers
to any of my math questions. She would either immediately answer a math question with a precise answer,
or give me the name of the best person to contact.
  
During her stay at IHES, Yvonne wrote some twenty-six research papers, two highest-level scientific books,
\cite{Choquet-Bruhat:2009xil} and \cite{Choquet-Bruhat:2014okh}, and her autobiography \cite{Ladymath}\footnote{for it, I had 
originally suggested to Yvonne the title: \textsl{``Tribulations d'une math\'ematicienne
dans ce monde \'etrange"}. She liked the word \textit{``tribulations"}, but it was not retained in the final title. 
On the other hand, the  word \textit{``strange''} remained.}.

I had uncountable discussions with her about the content of all her books. 
She kindly asked me to write a chapter for her monumental, testamentary book \cite{Choquet-Bruhat:2009xil}
on the Belinsky-Khalatnikov-Lifshitz conjecture concerning the  behavior of generic solutions of Einstein's equations
near a spacelike singularity. I was happy to do so, and I tried to formulate, in an as mathematically precise
way as I could, the conjecture suggested in the pioneering work of
Belinsky, Khalatnikov and Lifshitz \cite{Belinsky:1970ew}, and refined in
the many physics works triggered by it. 

Let me note in this respect the not so well-known fact that the main results of \cite{Belinsky:1970ew}, 
and notably the coupled second-order nonlinear differential equations
for the three local scale factors $a(\tau), b(\tau), c(\tau)$, were first publicly presented by
Isaak Khalatnikov in January 1968 in a seminar at the Institut Henri Poincar\'e in Paris\footnote{Private communication to T. Damour by 
Isaak Khalatnikov, partly confirmed by Demetrios Christodoulou
(However, Demetrios has no memory of the content of Khalatnikov's talk because, during the talk, J.A. Wheeler 
received a telegram announcing the good news that Demetrios was
officially accepted as a student at Princeton University). I think that the story of this seminar
has been included by Isaak in his autobiography: Isaak M. Khalatnikov,
\textsl{``From the Atomic Bomb to the Landau Institute: Autobiography. Top Non-Secret"} (English Edition) 2012
Springer, 228 pp., ISBN 978-3-642-27560-9}. The audience comprised in particular,
John Archibald Wheeler and Demetrios Christodoulou. When Isaak presented the $a-b-c$ ODEs, Wheeler
made the public remark that this looked like a mechanical system where the Universe was described by the
three coordinates $a, b, c$. One year later (14 April 1969), Charles Misner published his famous Lagrangian
description of the  complex dynamics of Bianchi IX universes (called by him \textit{``the Mixmaster Universe"}), mentioning
at the end a private communication from Wheeler in which he \textit{``suggested that studies of singularities by 
Belinsky and Khalatnikov had also found alternating Kasner-like epochs but with a very simple description in terms 
of a related parameter ($u$)"}.

Her second scientific book \cite{Choquet-Bruhat:2014okh} led also to many interesting discussions with Yvonne,
who wanted to be kept abreast of the points of contact between General Relativity and experimental or observational
facts. Let us recall in this respect that Yvonne (whose father, Georges Bruhat, was a physicist, well known in France for his work on optics
and his textbooks) considered herself as a ``failed" physicist who constantly aimed at understanding the real universe through its 
theoretical physics description, by using, and perfecting, mathematical tools.

During Yvonne's stay at IHES, two conferences were organized in her honor: one in March 2004 for her 80$^{\textrm{th}}$ birthday, and
one in January 2014 for her 90$^{\textrm{th}}$ birthday. Another special moment happened on February 11, 2016:  Yvonne and a group of 
scientists of IHES (including me) followed the live announcement
(from Washington, DC, USA) of the discovery of the first gravitational wave signal by the two LIGO interferometers.
After the end of her regular visits at IHES,  a one-day conference 
was organized at IHES in December 2023 to celebrate her 100$^{\textrm{th}}$ birthday (unfortunately in her absence).

Let me finally mention that I had detailed technical discussions with Yvonne about one of her later research papers,
namely her  streamlined, complete proof (valid in arbitrary space dimension $n$, and using only spinors on some (oriented) spacelike section 
$\Sigma_n$) of the (strong) positive energy theorem ($ E \geq |{\bf P}|$) in General Relativity. She kindly offered me to co-sign her
paper. But I felt I had not significantly contributed to her proof, and I declined this honor, and this token of friendship.
  
\subsubsection*{Epilogue}
 
To end this tribute to the memory of Yvonne, let me mention one of my last discussions (by phone) with her.
One day in the spring, or early summer, of 2018 (at a time when she had essentially stopped visiting
regularly IHES), Yvonne called me on my cell phone (I remember that I was
walking in the countryside) to tell me the good news that the  Italian Society of General Relativity and Gravitation 
(SIGRAV) had contacted her to inform her that she had been awarded  the 2018 Amaldi Medal,  to be received in
person at the next SIGRAV National conference, due to take place in September  in Cagliari (Sardegna, Italy).
She was very happy to receive this distinction because she always had had close links with Italy and very
friendly relations with many Italian scientists. However, she told me that she did not feel she had the energy to go 
there in person (she was 95 !). I offered to collect the medal on her behalf, and to make a small presentation of her life work. 
So it happened. It gave me the opportunity to study in detail some of her  most important contributions to gravitational physics, 
which I greatly enjoyed.
[However, as I had to travel after the SIGRAV meeting to a summer school in Ravello, I had the  slight practical problem to be cautious in 
carrying all over Italy this very  valuable Amaldi medal, made of solid gold!]

\subsection*{James Isenberg}

I really got to know Yvonne in 1982, while we were strolling along Lake Tai in the city of Wuxi in China. Wuxi was one of the stops on the 
tour which accompanied the 1982 Marcel Grossmann Conference in Shanghai. There were about 40 people on that three city six day tour, and 
along with the wonderful scenery, that tour gave us lots of opportunity to get to know each other. I had actually met Yvonne very briefly 
almost 10 years earlier, when she came to give a lecture at Princeton University, where I was 
a junior undergraduate. I remember the lecture very well, because I was working on the conformal method for constructing and parameterizing 
solutions of the Einstein initial value constraint equations, and Yvonne was one of the pioneers in developing this method. After her 
lecture, my undergraduate advisor Jimmy York introduced me to 
Yvonne, but as a young student meeting one of the iconic figures in mathematical relativity, I didn’t think that I had much to say to her 
that would be interesting. I do remember trying to talk to Yvonne with my high school French, but I think she found it pretty inadequate, so 
we quickly reverted to English. Our conversation lasted only a couple of minutes.

I don’t think that Yvonne had remembered our 1973 conversation when we met again in China in 1982, but during that six day tour, with all 40 
of us in an exciting but relatively unfamiliar setting, there was quite a bit of time for many of us to get to know each other, and my 
opportunity to chat with Yvonne happened on that stroll 
around Lake Tai. I told Yvonne about how I had visited Paris on a high school excursion in 1967 and all the wonderful things that I had seen 
during that excursion – including seeing performances by Johnny Hallyday and Charles Aznavour – but I don’t think that Yvonne was that 
impressed, so I think our conversation moved from that on to 
our common interest at the time in the newly developed theory of supergravity. I knew that Yvonne was celebrated for her epic early 1950s 
proof that the Cauchy problem for Einstein’s gravitational field theory was well-posed, and I mentioned to her that my work with Jim Nester 
on supergravity had led me to wonder if the Cauchy problem for supergravity was also well-posed. That question really got our conversation 
going, because Yvonne had also been interested in supergravity, and we spent much of our stroll 
talking about that. One crucial feature of supergravity is that it includes Fermionic fields – spin 3/2 – which necessarily anti-commute. 
This feature results in the 
standard tools for proving the well-posedness of the Cauchy problem for a hyperbolic PDE system requiring significant modification. Yvonne 
and I discussed some of the 
modifications which would be necessary during that stroll, and our subsequent communications – which back then had to be carried out using 
regular mail – led to our first 
joint paper along with our collaborators David Bao and Phil Yasskin. I remember being extremely proud to publish a paper with Yvonne.

Our stroll in Wuxi led to a long period of friendship and collaboration with Yvonne. It was not long after our joint publication of the paper 
on supergravity that Yvonne invited me along with Vince Moncrief to spend nine months working with her at Paris VI (Jussieu). That was a 
wonderful time both scientifically as well as socially. The 
three of us met regularly, both in pairs and altogether. A lot of our discussions involved various special versions of the conformal method 
for solving the Einstein constraints, which was the topic of that first lecture which I heard Yvonne present in Princeton in 1973.

Five of the projects that I worked on with Yvonne focused on the conformal method. As Yvonne’s proof of the well-posedness of the Cauchy 
problem for Einstein’s (vacuum) theory of gravity shows (together with collaborative work with Bob Geroch), if one specifies an initial data 
set on a three dimensional manifold $S$, which includes a Riemannian metric $h$ and a symmetric tensor $k$ which satisfies the Einstein 
constraint equations, then there is a unique (up to spacetime diffeomorphism) maximal 
spacetime solution $(S\times I, g)$ of the full Einstein equations, with $g$ a Lorentzian  metric. As well, for this spacetime solution $g$, 
$h$ is the induced first fundamental form on $S$ and $k$ is the induced second fundamental form on $S$. The Einstein vacuum constraint 
equations take the form of an under-determined nonlinear set 
of four partial differential equations to be solved for $h$ and $k$. The Cauchy problem for Einstein’s equations is an especially effective 
and practical way to obtain spacetime solutions, both analytically and numerically. But it does rely on developing a systematic way to 
produce solutions of the constraint equations. This is the goal of the conformal method. The idea is to specify “seed data” on $S$ consisting 
of a conformal equivalence class $ [h] $, a divergence-free and trace-free symmetric tensor 
$m$, a function $T$, as well as a function $N$, and then use these seed data to construct a determined partial differential equation set – 
the ``conformal constraint equations” – to be solved for a vector field $W$ and a conformal factor $F$. Presuming this set of conformal 
constraint equations can be solved, the seed data together with $W$ and $F$, allow one to construct $h$ and $k$ which satisfy the constraint 
equations themselves.

Work done by Andr\'e Lichnerowicz, Yvonne, Jimmy York, Niall O’Murchadha, Vince Moncrief, David Maxwell, Daniel Pollack and myself as well as 
others shows that so long as the function $T$ – which corresponds to the mean curvature of the initial data set – remains constant, the conformal method works 
very well. All of these names of contributing researchers 
reflects the fact that the conformal method has been applied not just to the Einstein vacuum constraint equations on compact manifolds $S$, 
but also to the Einstein constraint equations coupled to various “matter fields”.The conformal method has also been applied for seed data 
which is asymptotically Euclidean, asymptotically hyperbolic as well as with other asymptotic conditions. Yvonne has played a significant 
role in developing the conformal method for many of these various cases, and I am very privileged to have worked with her on this research.

While some of my collaboration with Yvonne involved email, almost all of our work together was done face-to-face. I am very happy that this 
was the case, because we had the opportunity to meet in Paris, in Italy, in Oregon (at my farm), as well as a number of other places. The 
focus was usually on working together on research, but we also had a number of opportunities to enjoy each other’s company on “road trips” 
and I learned what a warm and caring and interesting person Yvonne was. I have wonderful 
memories of the time that we spent together in many locations.

While much of our collaboration involved the conformal method, there is another branch of research in general relativity on which I have 
worked extensively with Yvonne. This research involves determining the behavior of the gravitational field in a neighborhood of the Big Bang 
in solutions of Einstein’s equations. Many years ago, 
Belinskii, Khalatnikov and Lifshitz conjectured that in the neighborhood of the Big Bang singularity in solutions of Einstein’s equations, 
the gravitational field would exhibit “asymptotically velocity term dominated” (AVTD) behavior which means that with respect to a given 
coordinate system, the spatial derivative terms in the solutions would be dominated by the time derivative terms. The effective result would 
be that observers following time-like paths towards the singularity would see the gravitational 
field evolve much like spatially homogeneous solutions. A later conjecture of these three physicists was that the behavior would not be of 
this sort, but rather the observers approaching the singularity would see an infinite sequence of epochs of spatially homogeneous solution 
behavior (which has been labeled by Charles Misner as ``mixmaster” behavior). 

While neither of these conjectures is believed to hold for general vacuum cosmological solutions, there has been numerical as well as 
stability analysis evidence that AVTD 
behavior is found in solutions of the Einstein-scalar field and Einstein-stiff fluid field equations, as well as in solutions of the Einstein 
vacuum equations with 
significant symmetry. As well as the intrinsic interest in the presence of AVTD behavior in cosmological solutions, it has been found that if 
AVTD does exist in a family 
of solutions, then it is relatively straightforward to determine if that family of solutions satisfies the “strong cosmic censorship 
conjecture” of Penrose, which is one 
of the major outstanding questions in mathematical relativity.

Not long after Vince Moncrief and I spent those nine months in Paris with Yvonne, Vince and I were able to prove that the “polarized Gowdy” 
family of solutions do exhibit 
AVTD behavior, and following that Vince and I and Piotr Chrusciel were able to prove that strong cosmic censorship does hold for this very 
limited family. In subsequent 
years, Vince and I  regularly visited the IHES Institute in Bures-sur-Yvette just outside of Paris, and consequently had many conversations 
with Yvonne about various 
topics in mathematical relativity, including the issue of AVTD behavior in families of solutions of Einstein’s equations. While proving that 
this behavior exists in entire 
families of solutions is generally quite difficult, a method was developed by Alan Rendall and his collaborators to show that in various 
families of cosmological 
solutions, AVTD behavior occurs at least in an infinite dimensional subfamily of solutions. In our discussions at IHES, Vince and I were able 
to convince Yvonne to work 
with us on applying these techniques to determine if AVTD can be found in subfamilies of solutions with considerably less symmetry than the 
Gowdy solutions. We were 
successful in showing that a number of families of solutions do include infinite dimensional subfamilies with AVTD behavior in a neighborhood 
of the Big Bang. When I was 
last able to visit Yvonne at IHES, we were hoping to extend this analysis to wider families of vacuum solutions, as well as to Einstein-
scalar solutions with no symmetries.

Whenever Yvonne and I got together outside of France, without hesitation we spoke in English. But when we got together in Paris, Yvonne 
really wanted me to speak with her 
in French. I might have guessed that she would know from our earliest interactions that my French was not very good. Vince’s French was much 
better. I remember that he 
told me that he was able to develop very good French by watching French TV while we were together in Paris in 1986. I should have tried that. 
I was actually okay speaking 
French – although it was probably grammatically quite flawed – but I was not very good at understanding Parisian spoken French. I pretended 
to understand what was being 
said by periodically saying “d’accord”. Yvonne was always polite enough to not let me know that I was faking it.

Yvonne was a wonderful research collaborator and friend. I miss her very very much.

\subsection*{Richard Kerner}

Yvonne Choquet-Bruhat, my beloved Teacher and Friend, passed away in February 2025; still, the memory lingers on,

In this tribute to Yvonne's memory I will not praise her great scientific achievements, nor will I recall her extraordinary biography. 
Both subjects were given an exhaustive treatment in the article \cite{DCRK2022} written jointly with Demetrios Christodoulou.

I have no doubt that in the texts included in this issue many colleagues will praise Yvonne's everlasting contribution to 
mathematical aspects of Relativity and Field Theory, the Cauchy problem, partial differential equations, and her textbooks 
\cite{Analysis} and \cite{Choquet-Bruhat:2009xil}, which continue to be the best 
introductions to rigorous methods in Mathematical Physics, Differential Geometry and General Relativity{.

I decided to concentrate my text on my first encounter with my then future teacher and benefactor, whose
influence on my scientific itinerary and my destiny turned out to be beyond imagination. And to write a tribute not only to Yvonne, but also 
to France which I was discovering in the late sixties, a country that adopted me and became my new homeland, a country so different from
what it is in its present state, that most of the facts I will relate here may sound today as a daydream or a fairy tale. 

After getting my Diploma in Theoretical Physics at Warsaw University in May 1965, I was given a teaching position as an assistant at the 
Faculty of Mathematics and 
Physics of Warsaw University, starting October $1^{\rm st}$ of that year, and became a PhD student of Andrzej Trautman, who had been freshly 
named a Professor of 
Theoretical Physics when he was only 35 years old. During my first visit to Paris in September 1965 I met a young Parisian girl, we fell in 
love, Dominique visited 
me in Warsaw, then the next year her family invited me to France, where we spent a month in Brittany. Finally we got married in August 1967, 
and Dominique came to 
live in Warsaw. I worked a lot, and my first scientific paper written on a generalization of 
Kaluza-Klein theory suggested by my advisor Andrzej Trautman was ready in 1968. Trautman sent it to the \textsl{Annales de l'Institut Henri 
Poincar\'e}, a respectable French journal, where it was accepted for publication.

But the overall political climate was worsening, especially in $1967$ after the $6$-day war during
which Israel miraculously escaped annihilation facing the joint armies of its Arab neighbors. It was perceived as an utmost humiliation  
by the Soviet Bloc, Poland severed diplomatic relations with Israel, and official propaganda accused Polish Jews of a lack of loyalty towards 
their homeland, because of their commitment to Israel. When half a year later students' unrest exploded in Warsaw, Jews were blamed for that, 
too. Purges began, and many Jews were fired from their jobs. For many of them, who were so assimilated that they considered themselves as 
Poles, it was a painful wake-up. The authorities openly invited Jews to leave the country for Israel - and most of the Jews did not wait for 
a second invitation.

I decided to resign voluntarily from my University contract and to leave Poland in the only legal manner offered by the authorities, i.e., to 
emigrate to Israel. 
A person choosing that option was automatically stripped of his or her Polish nationality and given a travel document as a stateless person.
Along with others, we took the train to Vienna, and proceeded to Paris, where we met Dominique's mother, a widow living in a modest apartment 
in Saint-Ouen. 

Although I had no real expectations to get a job or a stipend to continue working on my PhD thesis, I consulted the Paris telephone directory 
displayed in all
phone booths, and found the phone numbers of Laurent Schwartz and Andr\'e Lichnerowicz, whose books on modern differential geometry
and mathematical physics I had read and admired in their Russian translations.
Luckily enough, both were at home when I called them. Lichnerowicz was aware of my article in the \textsl{Annales de l'IHP}, 
 and Schwartz invited me to come to his home the very next day. He was interested in politics and asked me many questions 
concerning the current situation in Poland. He asked me to leave my phone number - when I had none -
and my address. Lichnerowicz gave me an appointment the next week, on October 6.

Two days later, I received a \textit{``pneumatique''} from Laurent Schwartz who told me to call Madame Choquet-Bruhat who was 
looking for a teaching assistant. Her phone number and the address in the Parisian suburb of Antony followed. 

I immediately went to the phone booth down the street, and made a call. I presented myself and was invited to come to meet Yvonne 
at her home the next day. I reached Antony and found easily the two-store house at 16 Avenue d'Alembert;
I rang, the door opened almost immediately, and I saw Yvonne in person.
She led me through the entrance to the living room, we sat down on a comfortable sofa, 
and an informal examination began, testing my knowledge of distributions, Banach and Hilbert spaces,
Fourier transforms, partial differential equations and General Relativity. 
Apparently, Madame Choquet found my skills sufficient enough to ensure that I could conduct tutorials on distributions and partial 
differential equations.

She took a sheet of white paper -- I remember that it was not the standard A4 format, it was only half a page -
and wrote a few sentences with her ballpoint pen.
The letter was addressed to the administration of the Faculty of Sciences, and I had to bring it personally to 
an office in  Building ``C", on Quai Saint-Bernard, facing the Seine river. 

The next day I entered the Administration Building and presented myself to the person in 
charge of recruitments and the series of miracles continued. Having opened the envelope she read two short recommendations: firstly, asking 
to ensure my enrolment as a teaching assistant, and secondly, to make official my becoming her doctoral student - preparing for the degree 
\textit{``Doctorat d'{\'E}tat"}.

All this was understood and accepted automatically. The letter I submitted was written on a sheet of white paper, without any official header 
or stamp. The hand-written signature was good enough to put into movement the well-oiled University Administration. A gently smiling lady 
gave me two forms to be filled in: the first was the work contract, and the second was my admission as a doctoral student, naming Professor 
Choquet-Bruhat as my official advisor. 

In the first months of 1970 Sorbonne University was divided into 13 independent universities, the former \textit{Facult\'e des Sciences} was 
given the entire campus built in place of the ancient wine caves \textit{``Halles aux Vins''}. It became the University Paris-VI, which very 
soon split into Paris-VI and Paris-VII. 
Yvonne Choquet-Bruhat, a full professor there since 1960, became the director of the \textit{Laboratoire de M\'ecanique Relativiste}, a part 
of the newly created Department of Mechanics of Paris-VI. 
   
In spite of being one of the youngest full professors, and the only woman with this title, Yvonne was respected as a renowned scientist by 
her peers. As a group leader, her authority was undisputable, although she never raised her voice, and behaved with a rare modesty and 
simplicity. She cared about the members of her group, especially the PhD students, ensuring as often as possible the participation in 
scientific meetings in France and abroad.
Yvonne had also many foreign visitors, some of them quite famous, to a great benefit of all. Her leadership qualities were obvious and 
appreciated by her team and by the University and CNRS authorities alike.

Without being aware, I was given a masterclass not only in research, but also in the art of teaching and taking care of Diploma and PhD 
students. These were a few lessons which shaped my attitude for the decades to come: 

\vskip 0.1cm
- To share my own research, especially the current work, with my students, inviting them to develop it further;
\vskip 0.1cm
- To encourage as often as possible participations in scientific meetings and workshops, meeting new people and learning new topics.

\vskip 0.1cm
Later, when my relationship with Yvonne grew into close friendship, I often presented my PhD students and post-doctoral visitors to her , 
asking for advice, and most importantly, to give my students an opportunity to meet a great scientist, a legend, and a role model.  

My infinite gratitude towards Yvonne is hard to express, and it would be very difficult to return back all that I owe her. So the only way to 
pay my debt is to bestow on my own students and pupils as much care as I received during my first years under her guidance and caring 
attention.

\subsection*{Sergiu Klainerman}

I met Yvonne on my first visit to Paris after finishing my PhD, probably in the summer of 1980. As she describes herself in her 
\textsl{“M\'emoires”}, I came to her office to 
ask her opinion whether my recent global existence result for quasilinear wave equations in space dimensions  $\ge 6$  could be of physical 
interest. This was 
mostly a pretext to meet her; her name loomed large in my imagination in view of her foundational results in General Relativity (GR), 
especially her famous local 
existence result for the Einstein vacuum equations (EVE). That result was in fact based on the observation that the EVE reduces, in wave (or 
harmonic) coordinates, 
to a system of quasilinear hyperbolic equations, similar to the type I have treated in my work. Though ignorant of GR when I first wrote my 
PhD thesis, I became 
interested in the subject during my stay in Berkeley as a postdoc, largely due to the influence of S. T. Yau who made me aware that my work 
could be relevant to the 
problem of the nonlinear stability of the Minkowski space.

My result had however a serious defect: it was restricted to high space dimensions, and as such it was inapplicable, in principle, to the 
physical case of dimension $n = 3$. Nevertheless, Yvonne, liked my result to the point that she could not resist the competitive temptation 
to point out that in such high dimensions she could 
have also proved a similar result. This was true, of course, but for me that remark was actually flattering. More importantly she mentioned 
to me that Einstein conjectured, at the time of her visit to the IAS in 1951-1952, that the Minkowski space is non-linearly unstable in the 
physical case of space dimension 3; this 
fact is mentioned in her autobiography \cite{Ladymath}. I was not too bothered by this crucial, negative, piece of 
information, since, at that time, I would have been happy to improve my result to all dimensions $ \ge 4$. Interestingly, as I 
learned later, in one of her less 
known papers (\textit{``Un th\'eor\`eme d’instabilit\'e pour certaines \'equations hyperboliques non lin\'eaires”}),
Yvonne seemed to confirm Einstein’s conjecture by identifying a possible obstruction  (the failure to 
satisfy the null condition). This short paper played a very important psychological role in my later work with Demetrios 
Christodoulou on the nonlinear stability 
of the 3+1 Minkowski space, as it meant that we had to avoid wave coordinates and rely instead on a more covariant approach.

During the same visit, Yvonne told me that her brilliant young collaborator Demetrios Christodoulou was also interested in the 
stability of Minkowski space and 
mentioned to me that he planned to visit the Courant Institute on a Humboldt grant.  This was great news to me as I was going to 
go to Courant too, in the fall of 
1980, as an assistant professor. In fact I did actually meet Demetrios earlier, during a short visit by him at Berkeley, where I 
spent two years (1978-1980) as a 
postdoc. Demetrios told me at the time that he was trying to prove the stability of Minkowski space using the conformal 
compactification method. This very clever 
method, first introduced by Roger Penrose as a heuristic tool to derive the decay rates of the gravitational field at null 
infinity, had been rigorously used by 
him, in collaboration with Yvonne, to prove the first global existence result for Yang-Mills-Higgs type equations. It turned out 
that the method was not quite 
applicable to the Einstein equations, because of the non-triviality of the ADM mass, yet the method turned out to be very 
influential in many other situations.

Demetrios’ visit at the Courant Institute (1981-1983) started indeed a very fruitful collaboration that led ultimately to a proof 
of the global nonlinear stability 
of the Minkowski space,  in  the physical  case of space  dimension $3$, thereby disproving, among other things, Einstein’s 
conjecture.

It is hard to overstate the importance of Yvonne’s work. She, together with a very select few others,  such as Leray and 
Lichnerowicz, was among the first 
mathematicians in the world to concentrate their attention to the great mathematical challenges of the general theory of 
relativity. At the time when she wrote her 
famous \textsl{Acta Mathematica} paper \cite{YCB52}, it was still 
debatable in the physics community if the Einstein field equations were hyperbolic. Her work, which treats the initial value 
problem for the Einstein field 
equations in full generality, can be rightly considered as the starting point of the systematic mathematical study of the problem 
of evolution in GR. A later 
extension of her local (in time) existence result, obtained in collaboration with Robert Geroch, associates to any initial data 
set which satisfies the Einstein 
constraint equations a unique maximal, future, global hyperbolic development, a crucial concept which provides the natural 
framework for the global study of the problem of evolution in GR.  

Yvonne made many other decisive  contributions to mathematical GR  including the following: 1)~on the constraint equation (work with 
Demetrios Christodoulou,  James York, James Isenberg and others), 2) global solutions to the Einstein vacuum equations with a $U (1)$ 
symmetry (in collaboration with  Vincent Moncrief), 3) global 
solutions for the Einstein vacuum equations in higher dimensions (in collaboration with Piotr Chrusciel and Julien Loizelet), 4) 
characteristic Cauchy problem (in collaboration with Piotr Chrusciel and José-Maria Martin-Garcia) and has written  many influential  books  
such as \textsl{``Analysis, Manifolds and Physics''} and 
the remarkable monograph \textsl{``General Relativity and the Einstein Equations''}.

I have met Yvonne many more times through the years, and it was always a great pleasure to talk to her, not only about mathematics and 
physics, but about any subject concerning our “strange universe”. Her wisdom, warmth, broad interests, generosity and intellectual honesty 
are something I will always remember about her.

\subsection*{Philippe LeFloch}

Around the year 2000, as I launched the \textit{S\'eminaire de Relativit\'e Math\'ematique} at Universit\'e Pierre et Marie Curie (now 
Sorbonne Universit\'e), 
Yvonne Choquet-Bruhat was very encouraging and  immediately showed great interest to participate. She never missed our monthly meetings, 
which sometimes also met at Observatoire de Paris-Meudon (with \'Eric Gourgouhlon's group) or Institut d'Astrophysique de Paris (with Luc 
Blanchet's group). This seminar was also attended by Jean-Pierre Bourguignon and Thibault Damour, Yvonne's colleagues from IH\'ES. 

I remember many exciting scientific conversations together, as Yvonne always had some precise responses to my questions and immediately 
pointed out precise references, at a time where I was excited to learn more about General Relativity. Her vast knowledge was truly impressive 
while she remained very friendly and accessible.

We also exchanged occasional emails -- always thoughtful, always courteous. In 2015, to mark the centennial of general relativity, I 
organized a trimester of activities at the Institut Henri Poincar\'e. Yvonne attended our main event {\it ``General Relativity: a celebration 
of the 100$^{\textrm{th}}$ anniversary''} (IHP, Paris, November 16 to 20, 2015), but told me she preferred not to speak. Her reply during the 
preparation of this event was friendly, as always: {\it ``Cher Philippe,
Merci pour votre message. J'ai renonc\'e au travail math\'ematique et je suis souvent en province chez l'un ou l'autre de mes enfants, ce qui 
m'empêche de participer comme je le voudrais aux conf\'erences que vous organisez. J'espère pouvoir le faire davantage en novembre. Cependant 
je d\'ecline l'honneur que vous me proposez d'une courte pr\'esentation historique, il faudrait un talent que je n'ai pas pour sortir des 
banalit\'es \'evidentes. Bien amicalement, Yvonne"}\footnote{\it Dear Philippe, Thank you for your message. I've stepped away from 
mathematical work and am often out of town, staying with one or another of my children, which unfortunately keeps me from attending the 
conferences you organize as I would like. I hope to be able to take part more 
in November. As for the kind invitation to give a brief historical presentation, I must respectfully decline---it would take a talent I do 
not possess to avoid saying only the most obvious things. With warm regards, Yvonne.}.

We happened to live in the same city, Antony, near the Parc de Sceaux. Yvonne insisted on staying in her nice but large house, and continued 
to take the RER B line to IH\'ES or to Paris, long after most people of her age would have chosen easier options. When she moved away from 
Antony, I would sometimes hear news of her from our neighbour, who was her long-time friend, Palmira. I will remember Yvonne as someone 
remarkably accessible and truly inspiring.

\subsection*{David Maxwell}

I first met Yvonne Choquet-Bruhat in 2002 in Carg\`ese, Corsica at a summer school 
on mathematical relativity. I was a graduate student, two years into my PhD, and just beginning work on the first ideas that would comprise 
my dissertation.  Although Yvonne was not one of the primary lecturers, she was a celebrated presence at the school, titled in her honor: 
\textsl{``50 years of the Cauchy problem in General Relativity''}. I did not know it at the time, but this was 
my career's launching point and Yvonne would be a recurring figure in its early trajectory.

Sometime after the Carg\`ese school, I had a brainstorming session with my advisor, Dan Pollack, and his collaborator, Jim Isenberg, thinking 
about topics that might be suitable for me to work on. One promising idea concerned certain well-posedness results for wave maps and I set to 
work.  I had sketched out enough of a plan to know that the ideas were viable around the time that I attended  a conference at Oberwolfach. 
While there, I took advantage of the well-appointed library 
for a literature search and discovered that I had been scooped, several years earlier, by Yvonne, in a short paper in a French journal that 
covered 80\% of what I had planned to do. Fifty years after her seminal work she was still going strong.

Undeterred by this setback, and inspired by a talk by Sergio Dain at the same Oberwolfach conference, I decided on a new research direction:  
construction of black hole initial data. As part of this work, I was also interested in determining the minimal regularity requirements for 
the Sobolev spaces used in these sorts of constructions. Hence I decided to work with metrics having just enough Sobolev regularity to ensure 
continuity, but not more.  This technical choice required an appendix containing the needed elliptic estimates, which generalized earlier
results by Yvonne together with Demetrios Christodoulou. When I was done with the paper, at the suggestion of my advisor, I sent a copy of 
the manuscript to Yvonne and received a brief, very fast reply: the work was nice, but she thought that the elliptic regularity would require 
\textit{``something more''} than what she and 
Christodoulou had done.  She was being gracious in telling me to look again carefully at those estimates. Sure enough, there was a gap: 
readily repaired, but certainly overlooked!

Looking back at the twenty years since that conference in Carg\`ese, the debt I owe Yvonne is clear. Nearly every publication of mine has 
cited her work. My primary research area overlaps with just one of her many interests, and sits on foundations she helped build. She was a 
model for me as a budding mathematician: prolific, 
rigorous, impactful, generous. I am just one of many students shaped by her influence, and her legacy extends well beyond her formal theorems.

\subsection*{Roger Penrose}

I have always had a great respect for the work of Yvonne Choquet-Bruhat as it is the kind of thing which I am not good at doing myself but 
which I depend upon 
crucially in my own work. She had a remarkable skill in developing important rigorous results concerning the implications of systems of 
differential equations and what one can say as the precise implications of such systems. I find it important in the work that I have done to 
be able to depend upon such definitive conclusions as applied to the implications of general relativity. In addition, I always found Yvonne 
to be charming, and helpful to the work that I have been able 
to proceed with myself, most specifically the technical conclusion that I arrived at in the 1965 paper that eventually won me a Nobel Prize. 

\subsection*{Dan Pollack}

As a mathematics graduate student in the late 1980’s I had a clear but naïve interest in general relativity. This was deepened when Richard 
Schoen, who would become my doctoral advisor, offered a course on the subject during my second year, soon after his return to Stanford 
University. 

Following a general introduction, the course focused on geometric and elliptic problems in relativity. The foundational 1952 work of Yvonne 
Choquet-Bruhat showing that there was a hyperbolic system of 
equations behind the Einstein equations and proving the existence and local uniqueness of solutions, and her subsequent work, published in 
1969 with Robert Geroch, which used this underlying hyperbolic structure to establish global well-posedness of the Cauchy problem for 
Einstein equations, was of course explained but not explored in depth in the course.  

My initial areas of mathematical research were strictly geometric and elliptic, focusing on problems related to scalar curvature and surfaces 
of constant mean curvature (CMC) in 3-space. 

I began to pursue my interest in general relativity thanks to Jim Isenberg, who was in the audience at a talk I gave at an AMS meeting 
at UC Davis on gluing constructions for CMC surfaces. Over the next few years my work turned towards developing gluing constructions for the 
Einstein constraint equations.  This led to a much deeper personal understanding of the enormous mathematical contributions of Yvonne 
Choquet-Bruhat. 

I first met Yvonne at the Summer 2002 conference on \textsl{``50 years of the Cauchy Problem in General Relativity”}, in Carg\`ese, Corsica. 
The Corsica meeting was of course a celebration of her 
seminal 1952 paper and everything that was enabled by that work. It was a treat for me, who at that point had just made a single contribution 
to the field, to attend that celebratory conference. I had by then come to recognize how nearly everything I knew about mathematical general 
relativity, beyond the Einstein equations themselves and their special solutions, rested squarely on her broad mathematical shoulders.  

Yvonne and I got to know each other and interact more meaningfully during the program on \textit{``Global programs in mathematical 
relativity”} at the Isaac Newton Institute of Mathematical Sciences in Cambridge, UK in Autumn 2005. It was there that we began our enjoyable 
collaboration on the Einstein-scalar field constraint 
equations which led to the three papers we wrote with Jim Isenberg. The (incorrect) assumption prior to then was that most results which were 
established for the vacuum constraint equations could easily be adapted to the constraint equations when minimally coupled to other non-
gravitational fields. 

It was David Maxwell, my former PhD student, who first observed that even the simple Einstein-scalar field system presented new and 
significant challenges not seen in the vacuum setting. It was such a great pleasure to explore this rich analytical playground with David, 
Jim and Yvonne. 

Yvonne was also very sweet with my family, who were with me in 
Cambridge. The occasion when she brought our young daughter a small white stuffed rabbit was memorable. Our time together in Cambridge was 
also an opportunity for me to benefit from Yvonne’s unique and deeply historically informed perspectives on the Einstein equations and 
analysis in general. I recall clearly a fascinating discussion we had on the history of “global hyperbolicity” and Jean Leray’s role in 
developing that notion.  This was a story that Yvonne was a direct participant 
in and it was fascinating to me to hear about this directly from her. In 2005, Yvonne was likely one of the very few people alive who could 
bring that history to life as a first hand participant.

I was so fortunate, as a relatively junior researcher, to fall into a research area that intersected directly with one of Yvonne’s main areas 
of interest and to have the opportunity to work with her. Her generosity and modesty were a very influential model for me. She was well into 
her 80’s when we started working together.  The curious sparkle and deep human vitality she embodied while discussing mathematics is 
certainly part of how she maintained such a continuous stream of 
impactful research over so many decades.  Her influence over the modern mathematical landscape, especially in mathematical physics and 
analysis, will have an everlasting impact. Fortunately, due to her many generous collaborations, I am not alone in carrying forward her 
influence and endeavoring to share her genuinely optimistic mathematical and human spirit with others.  

\subsection*{Tommaso Ruggeri}

In 1970, I had just graduated in Theoretical Physics, having taken several courses in Mathematics. I decided to attend the International 
Congress of Mathematicians 
(ICM) in Nice, France. That was the first time I saw and listened to a lecture by Yvonne Choquet-Bruhat. I do not recall the details of her 
talk, but I clearly 
remember that the room was packed. I was deeply struck by this charming and elegant lady, with a warm smile—who presented one of her recent 
results with remarkable 
clarity and confidence at the blackboard. At the time, I could never imagine that I would one day have the honor of collaborating with her.

A key figure in the French school of Mathematical Physics was André Lichnerowicz, one of Yvonne’s mentors and a close friend of Carlo 
Cattaneo, the great Italian mathematical physicist. Thanks to this connection, there were frequent exchanges between French and Italian 
scholars. Perhaps because of this, Yvonne spoke excellent Italian.

Cattaneo was deeply interested in nonlinear wave propagation and aimed to organize a CIME Summer School in Bressanone in 1980. Since I had 
begun working on problems in nonlinear wave propagation with Guy Boillat — who also had ties to Lichnerowicz — Cattaneo asked me in 1979 to 
help organize the school. However, shortly before it was set to begin, Cattaneo called to say he was unwell. As I was not yet a full 
professor (though I would soon become one), he asked one of his former students, 
Giorgio Ferrarese, to act as chairman and requested that I assist with the organization. 

A few days later, Cattaneo sadly passed away.

During the Bressanone school, I had the opportunity to meet Yvonne again, as she was one of the invited speakers. On that occasion, she 
presented her work on nonlinear asymptotic waves at high frequencies, one of her most significant contributions. In fact, Yvonne extended the 
method of high-frequency asymptotic waves — developed by Jean Leray, Peter Lax, and others for linear cases—to the nonlinear setting, aiming 
to apply this new theory to high-frequency gravitational waves.

I also gave a seminar at the school, presenting a result showing that any covariant quasi-linear hyperbolic system of balance laws compatible 
with a convex entropy law could be cast into symmetric form—provided one chooses a privileged field. Together with my first student, Alberto 
Strumia, we called this the "main field," and 
we even applied it to the case of a relativistic fluid. Yvonne greatly appreciated this result and asked me to publish it in the 
\textsl{Annales de l’Institut Henri Poincaré}.

That encounter marked the beginning of a mutual friendship and admiration. I was especially honored by Yvonne’s esteem for me.
On that occasion, together with Antonio Greco — a mutual friend of Yvonne and me, and a professor in Palermo — we invited her to visit our 
respective universities taking advantage of financial support of the CNR (Italian National Research Council). In 1982, she spent the first 
two weeks in Bologna with me, and the following two weeks in Palermo with Antonio.

When she arrived in Bologna, I was slightly anxious about what topics I could discuss with her. I knew general relativity but had never worked directly in the 
field. Having to speak scientifically with Yvonne made me feel a bit uneasy. To me, she was a legend. But as is often the case with great scientists, Yvonne 
adjusted the level of our discussion to mine and spoke with great simplicity. She explained to me that she had introduced a gauge to solve 
the Cauchy problem for the Yang-Mills equations on curved spacetime. 

I then asked her whether it would be possible to find an analogous gauge for the Einstein equations.
Our aim was to address the challenge of reformulating the 3+1 decomposition of Einstein's field equations into a hyperbolic system. By 
selecting an appropriate lapse function—corresponding to harmonic coordinates for time—and assuming a zero-shift vector, it is possible to 
combine the evolution equations with the constraint equations and cast the system into a strictly hyperbolic form. This is crucial, as it 
ensures the well-posedness of the Cauchy problem.

By demonstrating that the 3+1 system could be cast into a hyperbolic form, we facilitated the application of analytical and numerical methods 
to solve Einstein's equations—essential for simulations of phenomena such as black hole mergers and gravitational wave propagation. We first 
summarized this result in 1982 in the
\textsl{Comptes Rendus de l’Académie des Sciences} and published the full paper in 1983 in \textsl{Communications in Mathematical Physics}. 
The paper became quite well known. Later, Yvonne generalized our result with James York, removing the zero-shift condition.

Yvonne then invited me to spend a short period in Paris, which proved to be very fruitful. At the time, I was interested in constructing a 
hyperbolic theory of dissipative fluids in relativity. In Paris, Yvonne directed my attention to a paper by her former student and colleague 
Charles-Michel Marle, who had completed a thesis under Lichnerowicz on relativistic kinetic theory. Marle had extended Harold Grad’s 
classical moment method closure to relativity, obtaining a hyperbolic system - in 
contrast to the parabolic structure of the Eckart model. This paper became very important to me. 

Later, together with Ingo Müller and I-Shih Liu, we developed a relativistic macroscopic model based on the universal principles of 
continuum mechanics: the relativity principle, the entropy principle, and entropy convexity. Although our model was derived through a 
completely different procedure, it 
coincided with the kinetic closure obtained by Marle. This became the foundation of my ongoing work in the so-called Rational Extended 
Thermodynamics (RET). 

I later co-authored books on this subject with Ingo Müller (1993, 1998) and Masaru Sugiyama (2015, 2021).
The system we developed in RET is symmetric hyperbolic, and I proved that a condition known as the K-condition (due to Yasushi Shizuta and 
Shuichi Kawashima) allows for the existence of global smooth solutions, provided the initial data are sufficiently smooth. 
Yvonne appreciated my work and even asked me to contribute a chapter on this topic to her book on General Relativity.

I later discovered that Yvonne had suggested my name for foreigner membership in the French Academy. Although I was not ultimately selected, 
I was deeply honored that she considered me for such a distinction. Over my career, I’ve received several honors—including election to the 
\textsl{Accademia dei Lincei} at a young age—but I will never forget Yvonne’s gesture. Her esteem and friendship were the highest rewards I 
have ever received.

Although we did not publish any further papers together, we had many opportunities to meet. In Italy, we organize a biennial conference called
\textsl{Waves and Stability in Continuous Media} (WASCOM), now in its 23$^{\textrm{rd}}$ edition, in which Yvonne participated several times. 
These conferences often took place in beautiful locations in southern Italy, which she greatly enjoyed –- not only for the scientific 
discussions but also for the chance to explore the area during her free time. 

In addition, our friend Giorgio Ferrarese frequently organized small meetings on the Elba island, where I was pleased to 
meet many of Yvonne’s co-authors.

I was also honored to be invited to two conferences held in her honor, the most recent for her 90$^{\textrm{th}}$ birthday, organized by our 
mutual friend Thibault Damour. That was the last time I saw Yvonne. Since then, I have continued to send her New Year’s greetings each year. 
These past years, however, it was her daughter Geneviève and her son Daniel who replied.

Even though the mind can accept her passing as part of the natural order of life, the heart still struggles with her absence. She was an 
extraordinary woman who will be remembered in the history of Mathematical Physics -- a kind, humble woman with a sweet smile. I feel truly 
fortunate to have had the great privilege of being not only her co-author, but above all, her sincere and lifelong friend.

\subsection*{J\'er\'emie Szeftel}

Although I am too young to have known Yvonne Choquet-Bruhat personally, I had the privilege of meeting her on a few occasions at conferences. 
From these interactions, I remember her as a gracious person with a sharp mind and great humility. 

That said, I am very familiar with her work, as Yvonne Choquet-Bruhat is a prominent figure in the field, known for her foundational 
contributions to the study of the evolution problem in general relativity -- that is, the study of the dynamics of Einstein equations -- 
marked by the following three seminal breakthroughs:
\begin{enumerate}
\item 
First, as a PhD student, she proved the local well-posedness of Einstein equations in her celebrated 1952 \textsl{Acta Mathematica} paper 
\cite{YCB52}. This was a remarkable achievement, as the theory of well-posedness for hyperbolic partial differential equations -- a broad 
class of evolution PDEs, including Einstein equations -- was still 
in its infancy at the time, having been pioneered by the works of Juliusz Schauder, Jean Leray, and Sergue\"i Sobolev. 

Einstein equations posed a particularly challenging case, being quasilinear, subject to general covariance, and requiring the establishment 
of the propagation of the constraint equations satisfied by initial data sets.
 
\item 
Then, in her landmark 1969 \textsl{Communications in Mathematical Physics} paper \cite{CBRG} co-authored with Robert Geroch, she tackled the 
global evolution problem for solutions to Einstein equations. Specifically, she demonstrated that her local solutions could be uniquely 
extended as maximal globally hyperbolic developments. 

This work was again at the forefront of research, as the concept of global hyperbolicity for hyperbolic equations had only recently been 
introduced by Jean Leray. 

Other significant challenges included proving that the maximal developments could not be extended in any gauge, as well as addressing the 
issue of geometric uniqueness -- both of which arose due to the general covariance of the equations.

\item Finally, in a series of influential works spanning from the 1960s to the 1980s, she developed a framework -- still in use today -- for 
generating as many solutions as possible to the constraint equations. This framework allowed for the construction of a large number of 
initial data sets for Einstein equations, each leading to 
corresponding maximal globally hyperbolic developments in view of her earlier result with Robert Geroch.
\end{enumerate}

In addition to the foundational contributions mentioned above, Yvonne Choquet-Bruhat made several other deeply influential discoveries that 
continue to have far-reaching implications today. Two examples that immediately come to mind are:
\begin{itemize}
\item her work \cite{FB56} on the $1+3$ formulation of general relativity, six years prior to the celebrated paper by Richard Arnowitt, 
Stanley Deser, and Charles Misner, and her joint work with Tommaso Ruggeri \cite{CBTR} on the corresponding hyperbolicity of such 
formulation, which had a profound influence on later formulations widely used in numerical relativity;

\item her work \cite{YCB69} on the construction of high-frequency gravitational waves satisfying Einstein equations which serves as a 
precursor to current developments surrounding the Burnett conjecture on high-frequency limits of solutions to Einstein vacuum equations.
\end{itemize}

Finally, Yvonne Choquet-Bruhat continues to exert a profound influence through her numerous textbooks. In particular, her monumental 2009 
monograph \textit{General Relativity and the Einstein Equations} \cite{Choquet-Bruhat:2009xil} remains a rich source of inspiration, offering 
countless research directions for younger generations. 

While the field of mathematical general relativity -- and mathematics more broadly -- has lost one of its great figures, the scientific 
legacy of Yvonne Choquet-Bruhat is very much alive, and her work continues to inspire current and future generations of scientists, reaching 
far beyond those who had the privilege of knowing her personally.

\end{document}